\documentclass[aps, twocolumn, showpacs]{revtex4}
\begin{document}
\title{Electric Charge, Fractional Spin and Flux}
\author{S. C. Tiwari \\
Institute of Natural Philosophy \\
1 Kusum Kutir, Mahamanapuri \\
Varanasi 221005, India }
\begin{abstract}
Ratio of electron charge radius and Compton wavelength of electron is known to be equal to the dimensionless electromagnetic coupling constant $e^2 /\hbar c$. It is pointed out that the coupling constant has two alternative interpretations: as a ratio of two angular momenta since Planck constant has the dimension of angular momentum, and two flux quanta $e$ and $hc/e$. We argue that it has deep physical significance such that the electronic charge becomes flus itself and at a fundamental level fractional spin of quantized vortex.  A unified perspective of the three interpretations of the coupling constant is presented invoking the new interpretation of the magnetic moment of the electron comprising of three terms. A critical discussion on the past attempts to give fundamental importance to magnetism and flux quantum is given and the implication on the unification quest of our ideas is outlined.

\end{abstract}
\maketitle

1. INTRODUCTION

Dirac's relativistic equation of electron explains spin and magnetic moment of electron in a natural way \cite{1}, and re-interpretation
of negative energy states as anti-particle led to the prediction of positron discovered subsequently. It was a great triumph of Dirac's theory.
However as noted by Dirac \cite{2} it was in fact Weyl who first gave a definite statement on the mass of the anti-particle equal to the mass
of the electron while Dirac had originally identified that as proton. Spinning electron was first conceived by M. Abraham and a comprehensive
theory of electron having spin $h/4 \pi=\hbar /2$, where $h$ is Planck constant was given by Thomas \cite{3}. Thomas notes that Compton had suggested quantized spin in 1921 and Uhlenbeck and Goudsmit explained anomalous Zeeman effect using that suggestion. In contrast let us ask: What is electric charge? Is it not strange that even after more than a century of the discovery of the electron this fundamental question has not been raised? In a monograph devoted to electron \cite{4} I have reviewed numerous attempts to model electron and various equations of motion proposed in the literature to get rid off ambiguities and infinities. A question that has now become standard had been first asked by Dirac \cite{5}, namely the origin of charge quantization. It is instructive to quote Dirac: 'This smallest charge is known to exist experimentally and to have value $e$ given approximately by $h c/e^2$....... The theory of this paper ....... is found when worked out to give a connection between the smallest electric charge and the smallest magnetic pole.' Magnetic monopole and charge quantization have been of renewed interest in the light of the standard model of particle physics and beyond.

It is well known that one-particle Dirac equation is not satisfactory, and quantum electrodynamics (QED) also has foundational problems \cite{2}.
Here we mention two aspects of Dirac equation. Though spin half emerges beautifully, the magnetic moment has to be equal to the Bohr magneton
$\mu _B$ whereas the experimental value deviates from it by an anomalous term. The QED calculated value in the power of the fine structure constant
$\alpha =e^2/ \hbar c$ agrees extremely well with the empirical data and neglecting higher order terms is given by
\begin{equation}
\mu_e =\mu_B [1+\frac {\alpha}{2\pi} -0.328478444 {\frac {\alpha}{\pi}}^2 ]
\end{equation}
\begin{equation}
\mu _B =\frac {e \hbar}{2mc}
\end{equation}

It is only for the electron that anomalous part is a small correction, for proton that is also a spin half particle the magnetic moment predicted 
by Dirac equation would be $\mu _B m/m_p$, where $m_p$ is the mass of the proton. However it is in gross error from its actual value. A nice exposition on the magnetic moment of elementary particles with extensive list of
references can be found in \cite{6}. Thus Dirac equation cannot be accepted as a general equation for any spin half charged particle 

The second issue is that of nonlocality: though Dirac equation represents point particle a length scale of the order of Compton wavelength
$\lambda _c =\hbar /mc$ becomes necessary for physical interpretation. In some of the literature the Compton wavelength is taken to be $h/mc$. Dirac himself \cite{1} invoked high frequency oscillation (Schroedinger's zitterbewegung) in order to understand the motion of a free electron with the velocity of light predicted by the theory. It could be argued that localization of electron wavepacket in a spatial region of less than $\lambda _c$ would necessarily result into the interference between negative and positive energy states and the picture of one-particle breaks down. Curiously a length scale of the order of electron charge radius $r_e =e^2 /m c^2$ much smaller than the Compton wavelength appears at classical level in the electron theory; there is no explanation for this.

Physicists have pondered over the meaning of the dimensionless coupling constant $\alpha$, and speculated on the coincidence that the ratio of electron charge radius and Compton wavelength is equal to $\alpha$ relating it with structure of the electron. Attempts to build models of elementary particles using charge distribution over extended structures and also considering the role of magnetic moment have not succeeded. The advent of gauge field theories and the Standard Model of particle physics with remarkable successes has almost put a break on such efforts. Nevertheless there is no denying the fact that foundational issues on QED remain unresolved. In this short paper my aim is to put forward a new insight on the meaning of electric charge (in the next section) and briefly discuss its physical consequences which are likely to offer a radically new pathway to fundamental physics in the last section.

2. ELECTRIC CHARGE IS FLUX

A new ingredient in our model of electron \cite{4} is the proposition that electric charge is a manifestation of fractional spin of the order of
$e^2 /c$. In Sections 9.2 and 9.3 of the monograph \cite{4} the application of the tentative model of the electron was discussed in condendnsed matter systems, specially superconductivity and quantum Hall effect. In an otherwise positive review of the book, Post \cite{7} was critical of the fractional charge in Laughlin's theory of fractional quantum Hall effect apparently supported in my work. It has to be clearly stated that in \cite{4} electron is endowed with fractional spin not fractional charge, and this spin is interpreted to be the origin of charge $e$.

The argument behind this new interpretation came first from the observation that the fine structure constant being dimensionless immediately leads to the fact that $e^2 /c$ has the dimension of angular momentum since the Planck constant has this dimension. Does it have deep physical significance? Further support to this proposition was discussed in \cite{8} noting the remarkable fact that charge invariably occurs as $e^2$ in charge-field interaction if fields are expressed in the units factoring out the charge unit, and disappears for source free electromagnetic field equations. Visualizing pure spacetime fluid as fundamental entity the physically intuitive possible operations are those of spacetime translation and rotation, therefore it would be a great advancement towards unification if all interactions are ultimately caused by rotation and translation. A significant clue to it is offered by Equation (1) rewritten in the form
\begin{equation}
 \mu _e =\frac{e}{mc} [\frac{h}{4\pi} +\frac{e^{2}}{4\pi c}]
\end{equation}
Above expression could be interpreted to imply that electron has spin angular momentum of $e^2 /4\pi c$ besides the usual half spin. For notational convenience let us denote $e^2 /2\pi c$ by $f$. We have elucidated the significance of fractional spin  in \cite{9}, however there has not been further progress in this since linking the fractional spin with the charge has not been achieved in concrete form.

The new insight that makes the model of the electron complete to a great extent could be stated in the form of following hypotheses.

H1: Rest energy $mc^2$ of electron is purely rotational. Half of it is attributed to spin $\hbar /2$ and remaining half to fractional spin $f/2$.

H2: Electron is a composite structure of two vortices possessing magnetic flux quanta of $hc/e$ and $e$. The vortices have different core radii and the one with flux quantum $e$ behaves as a point votrex (or point charge); thus electric charge is nothing but a flux quantum.

We now give plausible arguments to justify the above hypotheses. It is interesting that in the theory of electron, spin angular momentum is not associated with energy, for example, Dirac on p.266 of \cite{1} states that,'The spin angular momentum does not give rise to any potential energy--'. It is only in the interaction that spin effect manifests, say in Zeeman splitting or the motion of electron in spherically symmetric potential. Similar situation prevails in the case of photon where its energy $h\nu$ is assumed purely kinetic, however we have recently argued that half of the energy of photon is due to spin angular momentum $\hbar$ of photon \cite{10}. It would be reasonable to associate rotational energy to electron. Interesting result follows based on hypothesis H1. Since fractional spin $f/2$ is postulated in addition to spin half we calculate rotational energy for the two separately in a simplified model of rotating disk using the formula $L^2 /2I$ where $L$ is angular momentum and $I$ is rotational inertia of the disk $MR^2 /2$ along the cylindrical axix of symmetry. Rotational energy for spin half would be equal to $\hbar ^2 /2ma^2$ and that for fractional spin it is given by $f^2 /2mb^2$ where $a$ and $b$ are the radii of the respective disks. Here mass for each disk is assumed to be half of the mass $m$. Equating each of these two with half of the rest energy we get
\begin{equation}
 a = \frac{\hbar}{mc} =\lambda_c
\end{equation}
\begin{equation}
 b = \frac{e^2}{2\pi mc^2} =\frac{r_e}{2\pi}
\end{equation}

Notice that the two lengths associated with the electron arise here in relation to the rotational energy; though we have used a simple picture this result is significant.

Next let us make further analysis of the fine structure constant: the fine structure constant could be considered as a ratio of $e$ and flux quantum $hc/e$ thus electric charge itself has the dimension of a flux. Hypothesis H2 becomes quite natural if this guess has physical content. Let us calculate the energy for two flux quanta recalling that a magnetic dipole placed in an external magnetic field has the energy $\mu B$. Since the hypothetical magnetic field is internal in the present case it is more appropriate to take half of this energy for electron magnetic moment. Multiplication by area of the disk renders the energy expression in the form of $\mu \Phi$ where $\Phi =B \times area$ is flux. For the assumed flux quanta $\Phi _1$ and $\Phi _2$ for electron we have
\begin{equation}
\frac{\mu_B \Phi_1}{2} =\frac{mc^2 \pi a^2}{2}
\end{equation}
\begin{equation}
 \frac{\mu_B \alpha \Phi_2}{4\pi} =\frac{mc^2 \pi b^2}{2}
\end{equation}
Substituting $a$ and $b$ calculated from the rotating disk model we arrive at 
\begin{equation}
 \Phi_1 =hc/e
\end{equation}
\begin{equation}
 \Phi_2 =e
\end{equation}

It is satisfying that the hypotheses H1 and H2 are consistent and that gives confidence in the present model. However it has to be pointed out that rotating disk model and 'magnetic energy' equal to $\mu B/2$ are approximations to what ultimately is suggested to be a vortex model. Regarding the factor of half in magnetic energy expression it is reasonable as magnetic field is internal to the electron and also this has been assumed in the Post's slender ring model of electron \cite{11}. For the sake of completeness if one assumes energy to be $\mu B$ then with the flux quanta of $hc/e$ and $e$ one can calculate the radii from Equations (6) and (7) respectively. These turn out to be somewhat odd looking $\sqrt{2} \lambda _c$ and $\sqrt{2} r_e/2\pi$, however for self-consistency it is possible to adjust the moment of inertia. Since the length scales (4) and (5) arise naturally in a new light we prefer the previous calculation.

The most precise experimental value of anomalous part in the magnetic moment of electron \cite{12} is equal to $1.159 652181 11(74)\times10^{-3}$ where (74) gives the 1-standard-deviation in the last digits. The expression (1) for $\mu _e$ on the otherhand originates from the perturbative QED calculations. Why should the individual terms in Eq.(1) be given fundamental physical significance? Recall that the leading term in the magnetic moment i. e. $\mu _B$, first predicted by Dirac equation is intimately related with intrinsic spin of electron, and in QED it is the lowest order term. Thus logically the magnetic moments at one-loop and higher orders could be attributed independent physical interpretation, and since the first term is related with spin we expect higher order terms to be in some way related with internal angular momentum or fractional spin of electron. We have already seen that fractional spin $f/2$ nicely relates with the electric charge, and gives a posteriori justification that one-loop level value has independent significance.

Historically the Stern-Gerlach experiment performed with molecular beams in a magnetic field did not measure directly the magnetic moment of electron; the experiment implied that the magnetic moment should be attributed to the whole of the atom. The idea that spin of electron does not correspond to a classical notion of spinning object but it is 'classically nondescribable two-valuedness of the quantum property' according to Pauli and the objections against the possibility of the measurement of spin magnetism of free electron led to interesting debate that continued till about mid-1970s \cite{6}, see also \cite{4}. Dehmelt at the University of Washington first measured the magnetic moment of free electrons, and later his group has achieved precision values. Amongst the papers from this group, in \cite{13} a brief discussion on some fundamental questions is of interest. Authors admit that a sort of quasiorbital radius of the order of Compton wavelength $\lambda _c$ for a point charge executing circular zitterbewegung could explain the Bohr magneton for electron, however absence of the magnetic moment interaction in high energy electron-electron and electron-positron collisions indicates that electron has to be viewed as a point particle with dimension less than $10^{-16}$cm. The dilemma faced between a point electron and extended structure is obvious from the remarks made in \cite{13}.

A miniature rotating golf ball model for spinning electron is certainly ruled out, however the present work brings together the three interpretations that are hidden in the expression of the fine structure constant: in terms of the ratios of two lengths (classical electron charge radius and Compton wavelength), two angular momenta (fractional spin and spin half) and two flux quanta ($e$ and $hc/e$). This must have a unifying picture. To achieve this let us consider an equivalent intrinsic spin angular momentum obtained from expression (1) for the magnetic moment comprising of three parts
\begin{equation}
 S_v=\frac{\hbar}{2} +\frac{f}{2} - 0.3284 f\alpha
\end{equation}
The magnetic moment expression is now translated to angular momentum, and is proposed to represent angular momenta of three vortices: quantized vortices in the rotating spacetime fluid or aether if one is not prejudiced against the usage of this term with the strengths (or circulation) $\Gamma_g$, $\Gamma_e$ and $\Gamma_w$ for the vortices, let us say central (C), orbiting (O) and tail (T) respectively. To fix the value of the strengths of the vortices one may divide $S_v$ by Planck constant and obtain dimensionless numbers. The sign of the circulation determines the sign of the charge: vortex-vortex repel and vortex-antivortex attract each other. In analogy to the electric charge $e$ embodied in the fractional spin $f/2$ we assign charges $g$ and $w$ respectively to the first and the last term of $S_v$ defined as follows

\begin{equation}
 \frac{g^2}{4\pi c} =\frac{\hbar}{2}
\end{equation}
\begin{equation}
 \frac{w^2}{4\pi c} =\frac{0.328\hbar \alpha^2}{2\pi}
\end{equation}

The strength of the vortex C is very large as compared to O and that of O relative to T. Therefore neglecting the vortex T for the timebeing, electron is envisaged as the vortex O rotating around C. Note that O has the spinning motion in a core radius of $b=r_e /2\pi$ besides the rotation around C. To the outside observer rotating vortex O appears as a source of point charge $e$: the Coulomb field is a manifestation of the time varying flux quantum $e$. Electron has two distinct internal configurations $O_-C_+$ and $O_-C_-$ where the suffix - or + denote vortex or antivortex. Other configurations $O_+C_+$ and $O_+C_-$ represent the positron.

It is necessary to clarify at this point that the preceding description deals with the internal structure of electron that lies in a 2+1 dimensional plane and this whole 3-vortex structure travels along the normal to the plane with light velocity. To put it in the context of Dirac's theory \cite{1} he obtains the time variation of the velocity of free electron having a component equal to the constant value $c^2 p/H$, where $p$ and $H$ are momentum and kinetic energy Hamiltonian respectively for a relativistic point particle, and an oscillatory part that has instanteneous value equal to the velocity of light. Dirac argues that the observed motion corresponds to a measurement of the average velocity over a time interval very large compared to the time scale of the oscillatory motion $h/2mc^2$. For a Dirac electron at rest there remains only oscillatory motion with light velocity. However in our model electron does not have a rest state in vacuum though the planar structure could possibly be confined in a three dimensional spatial region due to some sort of 'external potential' of a surrounding medium. Observed motion with velocity $v$ less than $c$ corresponds to the collision-limited average drift motion. The third length associated with electron, namely de Broglie wavelength becomes meaningful only in that case, and obviously it is not an intrinsic property of electron. The surrounding medium that we have invoked arises due to the presence of the multitude of the structures on the spacetime fluid or the substratum aether. The ubiquitous aether is not unphysical. Note that the whole of the universe is believed to be filled with some kind of fields in the modern theories both classical and quantum. Since we do not intend to identify the surrounding medium with specific notions such as quantum vacuum, microwave background radiation or curvature of spacetime having definite theoretical foundations, we prefer to call it aether.

A noteworthy standard result in the radiation theory of a charged particle is that for a uniform motion there is no radiation even if the charge moves with a relativistic velocity, however the field pattern surrounding the charge increases in the direction perpendicular to the direction of motion as the velocity increases, and becomes confined in the normal plane for the velocity approaching the velocity of light. For a charge at rest the field distribution is spherically symmetric. An accelerated charge radiates electromagnetic radiation: for a co-linear velocity and acceleration there is no radiation emitted along the forward direction of motion even if the charge is accelerated to the velocity near the light velocity. Planar vortex model of electron accounts well these features qualitatively in a rather obvious manner. Static electron here corresponds to a kind of revolving disk with approximately the dimension of $\lambda _c$ tracing out a spherical surface; it is due to the confinement caused by the surrounding medium. The disturbance caused by the vortex O propagates in the aether establishing isotropic Coulomb field when averaged out over the large time scales. Since free electron travels with velocity $c$, the accelerated motion actually represents the state when the density of the scatterers has been depleted along certain direction, and the average drift velocity increases. The physical mechanism of radiation would be the liberation of photons due to collision or intense disturbance caused by the vortex O in the plane of the electron.

In our model the circulation of central vortex C determines the spin state, up or down, of the electron and the charge $g$ associated with it is very strong at small distances. Spin polarized electrons $O_-C_+$ and $O_-C_-$ are markedly different from each other due the sign of the charge of C and the nature of vortex-vortex and vortex-antivortex interactions. Further we envisage a 2-vortex structure consisting of C and T only. The possible four states are proposed to be neutrinos and since the vortex O is absent these are `'electrically neutral'. Thus the two electron neutrinos and two muon neutrinos are identified with the for composite structures $C_-T_+, C_-T_-, C_+T_+, C_+T_-$.

3. DISCUSSION AND CONCLUSION

Reflecting on the past, the present ideas have to be seen as the continuation of the speculations on geometry being fundamental perceived by Riemann and Clifford, and Kelvin's vortex atom. In Kelvin's theory knotted vortex tubes of aether represented atoms accounting for their stability and variety \cite{14}. Would vortex dynamics and construction of knotted structures based on our model explain the existence of a large number of elementary particles and unify strong, weak and electromagnetic interactions? To address this question satisfactorily we must delineate what is distinctly new in our model, critically evaluate the past failed attempts that gave fundamental significance to magnetism or magnetic flux quantum, and ensure that the established physics is contained in the appropriate domain of the new theory.

We take up the last question first. Static Coulomb field and radiation find re-interpretation, at the same time the problem of infinity that plagues point field theory is eliminated. The asymmetry between the sources for electric and magnetic fields is not of fundamental nature as electric charge itself is a flux quantum: it is only due to the small value of this flux quantum $e$ as compared to $hc/e$, and the rotatory motion of the vortex O that for large distances the observed field is electric field. Assumption of the rotating flux $e$ as a point charge discarding the vortex C and the associated spin and flux quantum $hc/e$ we get the classical picture in which charges and currents are the sources of the electromagnetic fields. Spin of the electron does not play any role either in the description of the current flow or the Lorentz force law. Note that the magnetic moment arises as a secondary effect. Obviously none of the experimental facts would be violated in our approach. However rather than seeking magnetic monopole, the elusive object not observed till date, here we have electron as a composite particle consisting of electric-charge like flux quantum and magnetic-monopole like flux quantum. Further the two electrons with opposite spin have distinct internal structure, therefore the spin polarization of current carrying electrons should show up in new electromagnetic phenomena. Recent advances and interest in spin-polarized transport of electrons and spintronics have led to new effects and interpretations of electromagnetism \cite{15,16,17}. The present electron model offers the possibility for new insights in this field.

Though as early as 1917 the role of magnetic energy in the spinning electron model of Abraham was discussed \cite{3, 18}, in general magnetic field has been of lesser importance. Barut \cite{19} noted that,'It would have been strange if Nature provided magnetic forces just to be tiny corrections to the building principle of atoms ....'. In Barut's model the basic constituents of matter are assumed to be the stable particles proton, electron, neutrinos and photon, and the only binding force is that of electromagnetic origin. It is shown that magnetic forces between the stable particles become very strong at short distances; the strong interaction between hadrons is interpreted as a dynamical spin-spin and spin-orbit force. Lepton-hadron distinction is not of significance, and muon is visualized as a magnetic excitation of electron due to the interaction of the anomalous magnetic moment with its own field.

Schwinger in 1969 speculated on a magnetic model of matter \cite{20}. A new ingredient in his model is the modification of Maxwell equations incorporating Dirac's monopole, however postulating a new species of particle: dyon. Dyon is a dual charged particle possessing electric charge with coupling constant $\alpha$ and magnetic charge with coupling constant $4/\alpha$. A tentative theory of hadrons is outlined noting that the force between magnetic charges is superstrong in comparison with the strong nuclear force. Leptons are not composite though it is suggested that neutrinos could belong to both lepton and hadron families.

Jehle in a remarkable series of papers in 1970s not only highlighted certain fundamental questions in the historical perspective but also formulated a new approach to electromagnetism and elementary particle physics \cite{21}. The standard electromagnetism is built on electric charge and its dynamics. Jehle puts forward the hypothesis that quantized flux $hc/e$ is fundamental and the electricity and electric properties are derived from it. A closed flux loop is an elementary object from which a manifold of loopforms is constructed. Electron and muon are represented by a single loop. Topology of linked and knotted flux loops is used to interpret quarks and classification of elementary particles. It seems surprising that even after the advent of speculative superstring theory, Jehle's work has not received the attention that it deserves.

A more radical though tentaive idea is that of quantum cohomology due to Post \cite{11}. In his book Post makes two main contributions: a strong criticism of the orthodox Copenhagen interpretation of quantum mechanics, and an alternative topological approach for fundamental physics. Unfortunately excessive and repetitive emphasis on the first has obscured the novelty of the topological approach. Electromagnetism as metric-free theory, the recognition of flux quantum as de Rham period integral, and the significance of topological torsion in 4-dimensions comprise Post's quantum cohomology. To avoid likely confusion with the term quantum cohomology, it has to be emphasized that Post's idea is entirely different than quantum cohomology of superstring literature \cite{22}. Electric charge is fundamental in Post's theory, and a 3-dimensional period integral for spin angular momentum proposed by Kiehn \cite{23} is a new addition to the well known 1-dimensional Aharonov-Bohm flux integral and 2-dimensional Ampere-Gauss charge integral. Electron and muon are represented by a trefoil knot and a 'preliminary cohomological classification' for electron, muon, neutrinos, pions and photon is presented.

The question arises as to why these attempts did not succeed. Schwinger in his paper highlights the speculative character of his ideas and at one place remarks that,'However wide of the truth this hypothesis may be, it can serve to bring into better focus the nature of the quest for order and understanding that underlies the activity of the high-energy physicists'. It is possible that the ideas of Barut, Schwinger, Jehle and Post do not belong to the realm of the laws of Nature or have physical realization. This sort of conclusion would be rather premature since the mainstream physicists have not explored these ideas as vigorously as the most successful standard theories have been. Nevertheless let us have a critical look if there are weaknesses in these endeavours. One drawback common to them except that of the Post's work is that the new ideas were applied to the particle physics retaining the standard paradigm: the classification scheme based on so the called internal quantum numbers, conceptual framework of quantum field theory and quark models. Post argues for an alternative in which quark is not a legitimate object of physical reality; this, of course, would require tremendous effort to recast enormous knowledge in high energy physics in the alternative paradigm. Barut and Schwinger do not probe further if magnetism and magnetic charge could have deeper meaning than that following the Maxwell-Lorentz theory of electromagnetism. Jehle does take a step forward replacing the electric charge as fundamental to elementary flux loop as fundamental and also dispensing with magnetic monopole. I think there are two drawbacks in Jehle's approach that probably limited its scope. The derivation of electric field from quantized flux loop involves somewhat artificial introduction of a fraction of the quantized flux, denoted by F, since the loopform is assumed to spin at an angular velocity of $2mc^2/\hbar$ that corresponds to the Compton wavelength. Though quantized charge is explained due to the quantized flux without postulating magnetic monopole, there is no explanation as to what charge is. Secondly Jehle drifts away to construct quark model assuming them to be real physical objects.

Regarding quantum cohomology of Post it remains completely unexplored and ignored too. In my work \cite{10} a different idea than that of Post for topological torsion has been proposed for a new model of photon. Post treats flux as more fundamental than the magnetic moment, however electric charge is assumed fundamental elementary unit independent of spacetime provided by Nature. Has this rigid assumption on electric charge blocked deeper insights from quantum cohomology?

Now it becomes straightforward to state the new elements in our approach: in contrast to a single flux quantum $hc/e$ in Jehle's model we have two flux quanta such that electric charge itself is a quantum of flux, the classical concept of charge in conjunction with flux $hc/e$ could be used to derive the notion of magnetic moment but it is not fundamental, and finally the magnetic flux itself is a derived concept from the vorticity or the circulation of the spacetime fluid or aether vortex. The concept of electric charge proposed here is radically different than that of Post since spacetime rotation manifests as charge while according to Post charge is independent of spacetime. Neglecting the vortex T electron is a composite of two vortices or two flux quanta - it is akin to Schwinger's dyon. The concept of composite particles discussed in the literature on fractional quantum Hall effect should not be confused with our electron model. It has to be emphasized that in these theories the flux quantum of a vortex is created by the application of the external magnetic field on a 2-dimensional electron system; in the words of Stormer \cite{24} 'Electrons plus flux quanta can be viewed as new entities, which have come to be called composite particles, CPs'.

To conclude the paper we briefly outline the physical consequences of the new conceptual framework and suggest a fresh outlook on the 
unification quest. Fluid or hydrodynamical approach to electromagnetism in the nineteenth century, and to quantum theory in the 1920s is
well known. The present ideas in which electric charge is given a mechanical interpretation could stimulate revival of the fluid dynamical
 paradigm for fundamental physics: in \cite{8} the electromagnetic field tensor has been interpreted as angular momentum tensor
of photon fluid, the representation of sources in terms of flux integrals would render this description to a completion. Instead of a point
 charge what we have is a flux integral for electric charge, therefore the divergence problem will not arise. Quantized vortices
are best treated as topological objects making geometric and topological rendition of the electromagnetic phenomena quite natural.

Dirac theory of electron does not explain anomalous magnetic moment; theoretically it emerges as a QED effect. One could postulte
 additional term in the Dirac equation for electron under external electromagnetic field following Pauli to account for the
 anomalous magnetic moment. However the simplicity gets lost and the QED with this term is problemetic: though Pauli's term is Lorentz
 covariant and gauge invariant the resulting theory is not renormalizable. Could there be a possible modification to Dirac equation
that incorporates anomalous magnetic moment? Our model brings into sharp focus two length scales associated with the two flux
loops in the electron structure. Unlike Compton wavelength that emerges in the Dirac equation in various ways, the electron charge
radius is absent. We suggest the inclusion of fractional spin and this length scale into the Dirac equation such that the rest
energy $mc^2$ is treated as rotational energy could prove insightful. Since the sign of the electric charge is related with the circulation
of the vortex O, reformulation of Dirac theory in this perspective may ultimately resolve the issue of negative energy states. In an
 important work Gurtler and Hestenes  \cite{25} show that the consistency of Dirac and Schroedinger theories leads to the conclusion that
 the Schroedinger equation describes a particle in a spin eigenstate not a spinless one. Difference between charge current and momentum
 current in Dirac equation is also elucidated by the authors. The inconclusive ideas of Hestenes could be revisited in the light of
 present work and also that related with Weyl space \cite{26}.

If the present ideas have elements of physical reality the most profound implication would be on the nature of fundamental interactions.
The expression for $S_v$, Equation(10) consists of three terms and has a sort of universality: only fundamental constants $h, c, e$
appear in this expression. Electric charge is a quantized flux of the vortex O. Postulating tiny quantized vortices C, O, and T as
 fundamental constituents of the elementary particles the interpretation of the electric charge is extended to the new charges
$g, w$ given by Equations (11) and (12) respectively for the vortices C and T . The coupling constants in analogy to $\alpha$ are 
$2\pi$ and $0.656\alpha ^2$ for these charges. Since the strong coupling constant in quantum chromodynamics and Fermi's coupling constant
 for weak interactions are energy dependent, it is not sytraightforward to identify $g^2 /\hbar c$ and $w^2 /\hbar c$ with them, however
 the order of the strengths and the short range nature indicate that all the three fundamental interactions are embodied in $S_v$.

Electron is a composite structure of all the three vortices, thus confined in small spatial regions it could become a strongly interacting
 particle. Neutrinos are composite of C and T, and hence possess strong and weak interactions. Entangled links and knotted structures built
 from these basic building blocks represent the whole spectrum of the elementary particles. Simplicity and intrinsic unity of the present
 ideas, though speculative, invite attention as an alternative to the unification paradigm.

The library facility of Banaras Hindu University is acknowledged.

\end{document}